# Sensiverse: A dataset for ISAC study

Jiajin Luo[1], Baojian Zhou[1], Yang Yu[1], Ping Zhang[1], Xiaohui Peng[1], Jianglei Ma[1], Peiying Zhu[1], Jianmin Lu[1], Wen Tong[1]
[1] Huawei Technologies Co., Ltd
Corresponding author: Jiajin Luo (luojiajin@huawei.com)

*Abstract*—In order to address the lack of applicable channel models for ISAC research and evaluation, we release Sensiverse, a dataset that can be used for ISAC research. In this paper, we present the method of generating Sensiverse, including the acquisition and formatting of the 3D scene models, the generation of the channel data and associations with Tx/Rx deployment. The file structure and usage of the dataset are also described, and finally the use of the dataset is illustrated with examples through the evaluation of use cases such as 3D environment reconstruction and moving targets.

*Index Terms*—Sensiverse, ISAC, dataset, channel model, 3D environment reconstruction, moving targets detection.

## I. INTRODUCTION

THE primary purpose of wireless communication is to transmit information reliably, while one of main objective of wireless sensing is to obtain environmental information accurately. Although communication and sensing have different application scenarios and purposes, they both use radio waves as carriers and are related to the transmission or acquisition of information respectively. From this perspective, communication and sensing technologies have a natural basis for their integration. Additionally, with the breakthrough development of Artificial Intelligence (AI), there is a rapid increase in demand for data, and the collection of raw information from the physical world has also become increasingly important [1] [2]. In these contexts, the integration of communication and sensing has gradually attracted widespread attention from both the academic and industrial communities in recent years, with ITU-R approving the Integrated Sensing and Communication (ISAC) as one of the six major usage scenarios for IMT-2030 (6G) [3], and 3GPP SA1 [4] establishing SI to identify the application scenarios and requirements of ISAC for 5G-Advanced."

Currently, the industry is still exploring methodology to effectively evaluate the performance of ISAC systems [5] [6] [7]. In traditional cellular network (5G and earlier) research, wireless communication channel models have been widely used in link-level and system-level simulations, such as for evaluating KPIs like throughput, latency, BER (bit error rate), PER(packet error rate), among others. Both the industry and academia have done a lot of related work and constructed a relatively comprehensive evaluation methodology. However, due to the fact that sensing service requirements, such as accuracy and resolution of measurements, are different from communication services, the existing communication channel models [8] [9]cannot be applied to evaluate the sensing performance directly. In communication evaluation, channel models are used to simulate the impact of surrounding environments on the communication performance between transmitting and receiving nodes. Its evaluation metrics serve the throughput and reliability and timeliness of information transfer between transmitting and receiving nodes. It abstracts the environment into an information transmission pipeline and presents it in the time domain or frequency domain form. Since only the impact of environment on communication is concerned, the communication channel model does not require a precise modeling of the environmental information. Typically, a stochastic channel model with statistical characteristics is used for wireless communication standard development taking consideration of the diversity of the simulated environment and the simulation complexity. The stochastic model is not suitable to provide true value of sensing environment. It may also fail to accurately reflect the statistical characteristics of the sensing environment, making the stochastic channel model unsuitable for directly used for sensing.

In the view of the above issues, there are currently two primary types of channel models used to evaluate sensing: hybrid channel models and deterministic channel models.

The hybrid channel model typically uses a cloud of scattering points to represent the target to generate the deterministic component of the channel, while using a random component with statistical characteristics to simulate clutter [10]. Its advantage is the low computational complexity, which allows for the rapid generation of the corresponding channels. However, there is room for improvement, the deterministic scattering points and random component may not accurately reflect the true physical characteristics of real-world environments, especially in imaging scenarios where the clutter may be associated with the sensed environment's secondary echoes. For example, in certain SLAM(simultaneous localization and mapping) scenario, eliminating target clutter may be directly related to extracting "noise" information from the environment, so simply simulating the channel with scattering points and statistical characteristics may not effectively reflect the algorithm's performance and correctly evaluate the algorithm and design, more accurate deterministic models than scattering points may be needed. Based on deterministic calculations, such problems can be avoided to a certain extent [11] [12]. However, deterministic modeling calculations even in a single scene require a large amount of computing power, and in order to effectively evaluate the performance of the solution, it is usually necessary to simulate it in a large number of scenes.

To solve the above issues, we build a sensing channel dataset -- **Sensiverse**, which includes channel data in different scenarios and a large number of scenes and can bring the following benefits:

1) Building a multi-scenario and multi-mode sensing channel dataset supports applications such as moving target detection, 3D environment imaging and reconstruction, and integrated sensing and communication solution evaluation.

2) Avoiding the problem of evaluating results that do not converge to the real performance due to a limited number of evaluated channel samples mentioned earlier.

3) Mitigating the problem of high computational complexity. The dataset is generated offline and stored, which exchanges storage for computing power to reduce the computational resource consumption required by subsequent large-scale simulation evaluations.

4) The dataset can be used for calibration and training of hybrid channel models and for training and testing in machine learning for sensing.

In the generation of the dataset, we use raytracing to reduce the computational complexity of the deterministic EM(electromagnetic) simulation.

## II. Generation and structure of the dataset

As Fig. 1 illustrated, 3D map and Tx/Rx deployment information are fed into raytracing engine to calculate channels, and then the channels are converted to frequency domain.

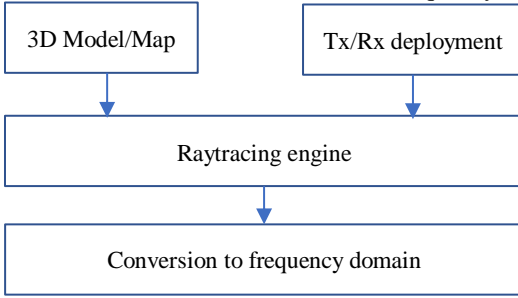

**Fig. 1.** The diagram of dataset generation.

Sensiverse consists of three parts, 3D maps, channel data and the example files as illustrated in Fig. 2

The 3D map section contains 3D map model files for cities. The channel data section is first divided into different folders according to the city, and then categorized into environmental sensing and target sensing in the city folders. Under each case, it is further divided into mono-static and bi-static sensing modes. Under each mode, it is classified by role(BS and UE) and finally stored according to frequency band. The example file section includes examples of environmental reconstruction and moving target tracking etc.

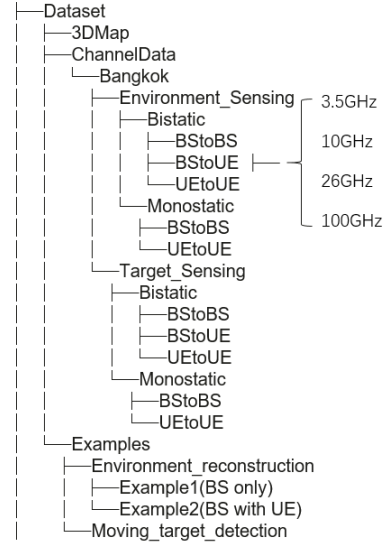

**Fig. 2.** The structure of Sensiverse.

### A. 3D Maps

As shown in Fig. 1, first, the generation of the dataset requires a 3D map model. We download 3D maps in the .osm format from OpenStreetMap and then convert it to .obj format. Because electromagnetic parameters and other information need to be configured according to the layers for ray tracing simulation in later stages, we performed layer aggregation on the obj file to form five layers: Building, Vegetation, Bridge, Water, Ground. In the first release version, we selected a total of 25 cities around the world as shown in Figure 3 and Table 1, where each city is represented by approximately 1-2 areas of about 4 square kilometers each. We will continue to add more map models to our dataset to support more application evaluations.

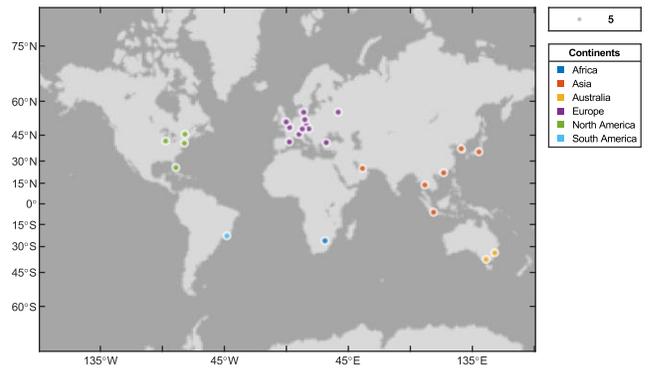

**Fig. 3.** The cities across the world.

TABLE I
Cities Across the world

| City | Continent |
|------|-----------|
| Johannesburg | Africa |
| Tokyo | Asia |
| Seoul | Asia |
| Hong Kong | Asia |
| Jakarta | Asia |
| Dubai | Asia |
| Bangkok | Asia |
| Sydney | Australia |
| Melbourne | Australia |
| Prague | Europe |
| Vienna | Europe |
| Paris | Europe |
| Milan | Europe |
| Moscow | Europe |
| London | Europe |
| Munich | Europe |
| Copenhagen | Europe |
| Berlin | Europe |
| Barcelona | Europe |
| Istanbul | Europe |
| New York | North America |
| Montréal | North America |
| Miami | North America |
| Chicago | North America |
| Rio de Janeiro | South America |

## B. Channel data: Composition and format

The outermost directory of the channel data section is the geographic location - city name, which corresponds to the city model in the 3D map directory. Because the channels for static environmental sensing and moving target sensing are different, the former only requires the deployment information of the map and transceivers besides 3D maps, while the latter also requires target status information that changes over time. Therefore, the city directory is further divided into environmental sensing and target sensing according to different applications. In addition, since sensing may work in mono-static or bi-static mode, the application directory is further divided into mono-static and bi-static sub-directories. Typically, the antenna array, capabilities, and mobility of base stations and UE are different, so they are further divided based on the different combination modes of node roles. Specifically, the mono-static mode is divided into BS to BS, UE to UE, while the bi-static mode is divided into the combination of BS to BS, BS to UE(or UE to BS), and UE to UE. After assigning the city, application, working mode, and role, the channel data is placed in a subdirectory named after the working frequency. Due to the convenience of using frequency domain channels for communication and sensing evaluation, we use frequency domain channels as the publishing format for our dataset.

The published channel files are saved in a file named H_freq.csv, consisting of three columns, representing the frequency, real and imaginary parts of the channel frequency response, and the unit of frequency is Hz.

TABLE II
FORMAT OF CHANNEL FREQUENCY RESPONSE

| Frequency | Real of H_f | Imag of H_f |
|---|---|---|
| ... | | |
| F_start+k*Δf | H(fk).real | H(fk).imag |
| ... | | |

Information about transmitter and receiver antenna posture and directional pattern is saved in a file named 'metainfo.txt'. The channel file 'H_freq.csv' and meta-information file 'metainfo.txt' are saved in a folder named after the relative position coordinate of the receiver and transmitter antennas. The folder name is like 'Tx_{x}_{y}_{z}_Rx_{x}_{y}_{z}', where {x}, {y}, and {z} represent the position coordinates of the Tx or Rx antenna elements. Users can choose the channel of different receiver or transmitter elements according to the position information in the filename based on their needs.

## III. USAGE AND EXAMPLES

### A. Real aperture sensing

In this section, the channel files selected from London_-0.13Lon_51.50Lat_0.020_0.020Side is taken as an example. Specifically, channel frequency response (CFR) from *Environment/Monostatic/BStoBS/10GHz* is chosen for real aperture imaging task. This folder gives 4096 CFR sub-folders from four monostatic Base Stations, and each Base Station is configured as 1Tx1024Rx. The layout of the scene is given in Fig. 4, where the white points represent the position of the four Tx. Meanwhile, the Tx/Rx antenna parameters can be loaded from *metainfo.txt* in each sub-folder.

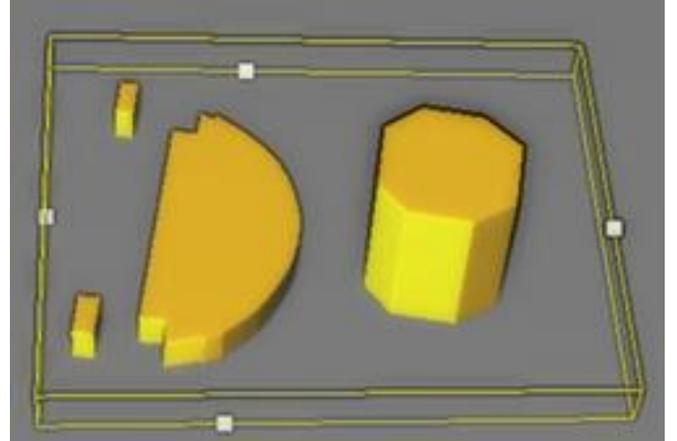

**Fig. 4.** Layout of real aperture imaging scenario.

To give an example of real aperture imaging task, Digital Beamforming (DBF) technique is taken to show the usage of the sensing channel dataset.

Firstly, Frequency Modulated Continuous Wave (FMCW) is selected as the transmitted signal s(t), where the carrier frequency is 10 GHz, band width (BW) is 399.96 MHz (9800 MHz to 10199.96 MHz from the first column of H_freq.csv), and the sampling frequency (fs) is 399.96 MHz, which is equal to the bandwidth. In according with *H_freq.csv*, the subcarrier spacing (SCS) $\Delta f$ is 60 kHz. The echo signal for each Rx can be depicted as

$$Y(f)=S(f)*H(f). \quad (1)$$

where $S(f)$ is the Fourier Transform of s(t) and $H(f)=Real(H\_f)+1j*Imag(H\_f)$.

Secondly, the element of each Rx is taken from Uniform Rectangle Array (URA). Therefore, the two-dimensional

Steering Vector a(θ,φ) can be calculated. The received signal R(θ,φ,t)after digital Beamforming can be treated as

$$R(θ,φ,t)=y(t)*a(θ,φ). \quad (2)$$

where θ depicts azimuth, and φ depicts elevation.

Lastly, the imaging result in each direction can be calculated by convolving the received signal with the conjugated time-reversed version of the transmitted signal s(t), i.e., by matched filter technique. The imaging fusion result from the four Base Station (BS) is shown in Fig. 5.

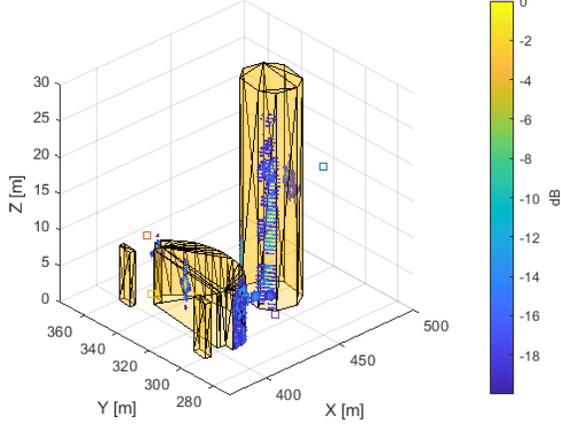

**Fig. 5.** Imaging fusion results from the four base Stations.

### B. Virtual aperture sensing

In this section, the channel files selected from Bangkok is taken as an example. As shown in Fig. 6, four BS are arranged around an elliptical building with a height of about 65 m. Directly below each BS, there is a UE moving along a straight line, and the height of the UE is about 1.5 m. Each BS forms a bi-static virtual aperture with the UE directly below it, where the UE serves as the Tx, the BS serves as Rx, and the elliptical building is the imaging target.

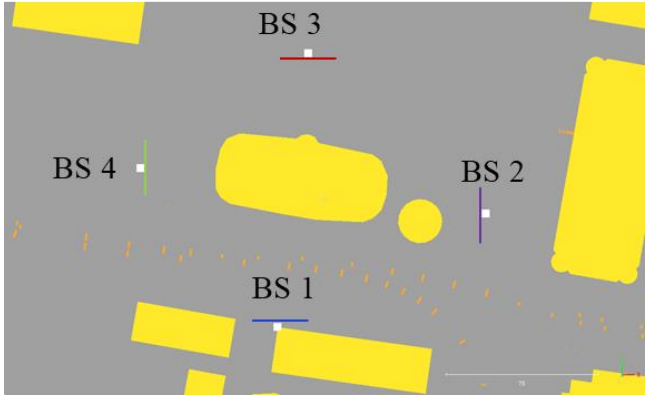

**Fig. 6.** Layout of virtual aperture imaging scenario.

Specifically, CFR from *Examples\Environment_reconstruction\Example2 (BS with UE)* is chosen for virtual aperture imaging task. This folder gives 48896 CFR sub-folders from four BS, and each virtual aperture is configured as 64Rx and191Tx.

Similar to real aperture sensing, firstly, the FMCW waveform is selected as the transmitted signal, where the carrier frequency is 100 GHz, BW is 1999.92 MHz. Then, Fourier transform of the transmitted signal is multiplied with CFR to form the echo signal. Finally, the back projection (BP) algorithm is used to process the echo signal to get the corresponding imaging results.

Fig. 7 shows imaging results obtained by four different virtual apertures, and Fig. 8 is a fusion result of the imaging results of the four virtual apertures. It can be seen from Fig. 8 that the imaging results have a relatively high resolution and can basically reflect a geometric shape of the elliptical building.

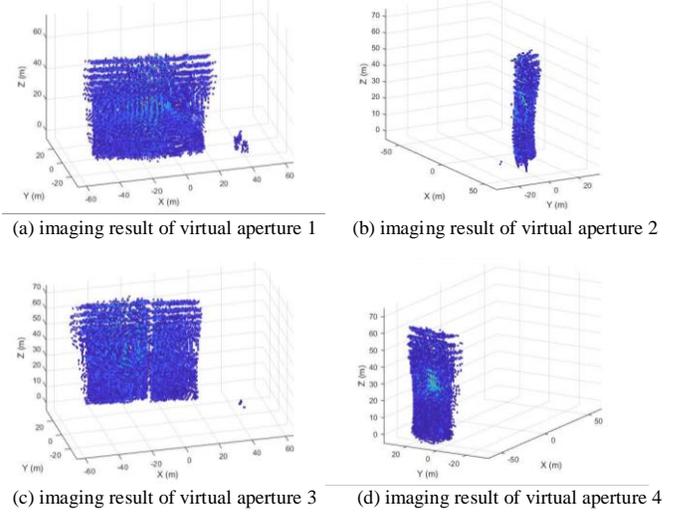

(a) imaging result of virtual aperture 1    (b) imaging result of virtual aperture 2

(c) imaging result of virtual aperture 3    (d) imaging result of virtual aperture 4

**Fig. 7.** Imaging results of four different virtual apertures.

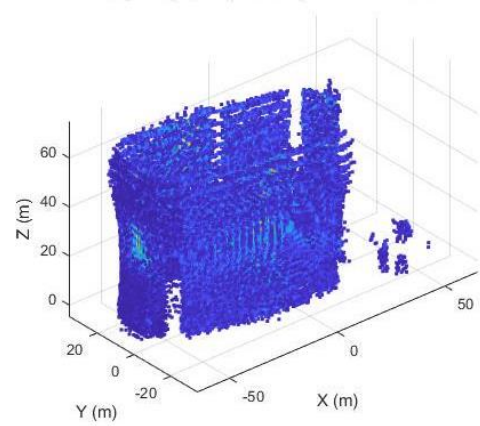

**Fig. 8.** Imaging results of four different virtual apertures.

### C. Network sensing for moving target sensing

In this section we take the city of Bangkok as an example as shown in Fig. 9, and its corresponding map file is 100.55Lon_13.75Lat_0.02Side.obj. We have selected a square area in this map and placed four stations around the square area. The positions of the four stations are (144.9,-90.6,10), (-67.52,-56.46,10), (-99.76,-227.16,10), and (97.49,-282.16,10), respectively. Each station is equipped with 1Tx and 256Rx. The Rx is a Uniform Rectangle Array (URA) of 16*16 antenna elements, and the minimum spacing between the antenna elements is λ/2, where λ is the wavelength of the RF signal. Using an 16*16 antenna array placed on the YZ plane with normal direction pointing in the direction of (1,0,0) as a reference, the antenna arrays on the four stations are rotated by [-135,-45,45,135] degrees around Z-axis so that the beam can cover the moving target. Each station operates in mono-static mode and can obtain the position information of the moving target through angle and range measurements.

The target motion consists of four straight-line trajectories, each with a length of 30 meters and containing 30 frames. The distance between adjacent targets in each frame is 1 meter, and each frame contains 64 snapshots with a spacing of $\lambda/4$ between adjacent snapshots. The starting and ending coordinates of the four trajectories are (5,-175,0)→(35,-175,0), (35,-175,0)→(35,-145,0), (35,-145,0)→(5,-145,0), and (5,-145,0)→(5,-175,0). The target used here is a cylinder with a radius and height of 3 meters. the Tx/Rx antenna and moving target information can be loaded from *metainfo.txt* in each sub-folder.

The channel response is provided in frequency domain H(f), with a center frequency of 10GHz and a bandwidth of 399.96 MHz (ranging from 9800 MHz to 10199.96 MHz from the first column of H_freq.csv), and a subcarrier spacing of 60KHz. The channel frequency response (CFR) is loaded from *Examples\Moving_target_detection* in this example.

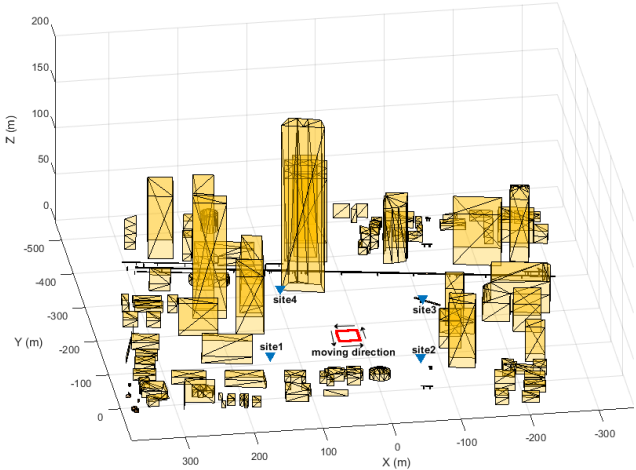

**Fig. 9.** Illustration of the scene for moving target sensing.

The simulation scenes have been explained above, and here we further provide the simulation signal processing flow. As shown in Fig. 10, the simulation process can be roughly divided into waveform generation, echo signals generation, obtaining RD (Range-Doppler) map, extracting target information, and calculating target position information based on angle and range information.

To be more specific, the waveform s(t) used here is FMCW signal, with a sampling rate of 399.96 MHz. The echo signal can be represented as Y(f)=S(f)*H(f). After obtaining the echo data, it is necessary to eliminate the influence of static clutter contributed from the environment through interference cancellation algorithms (e.g., ECA), and then further obtain RD map through beamforming, matched filtering. Based on the RD map results, CFAR operation can be performed to obtain range and angle information, and finally obtain the target position information. Fig. 11 shows the results of target positions detected by four stations.

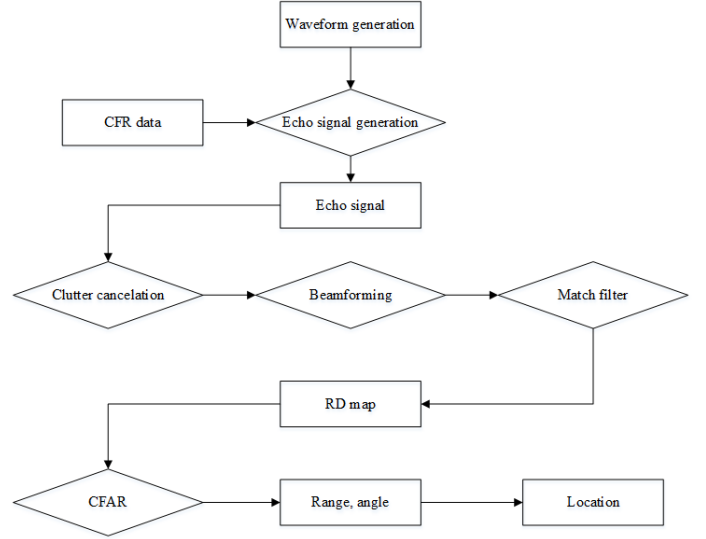

**Fig. 10.** Signal processing flow chart.

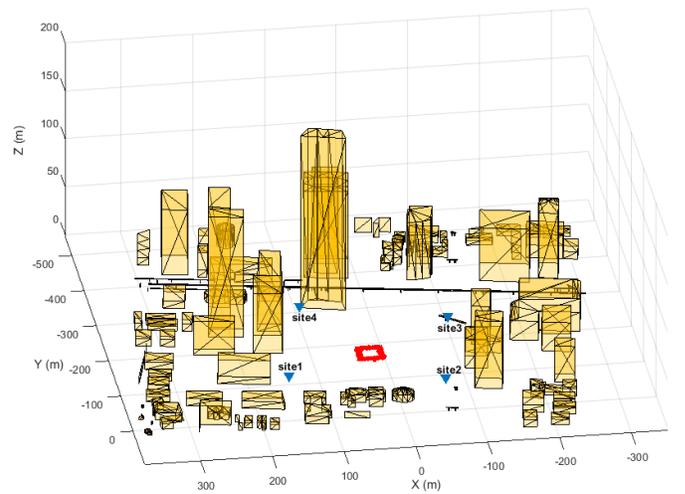

**Fig. 11.** Moving target detection results.

## IV. CONCLUSION

In this paper, we offer the dataset **Sensiverse** as a tool for future ISAC research and evaluation. The dataset can support the evaluation of a variety of sensing scenarios and use cases, including environment reconstruction and target tracking. The generation method, saving format and usage of the dataset are also introduced. We will continue to update and enrich its content in the future, and welcome feedback and suggestions from industry and academia to participate in expanding the dataset together.

REFERENCES

[1] Huawei, "Integrated Sensing and Communication (ISAC) — From Concept to Practice," [Online]. Available: https://www.huawei.com/en/huaweitech/future-technologies/integrated-sensing-communication-concept-practice.

[2] Hexa-X, "Localisation and sensing use cases and gap analysis," [Online]. Available: https://hexa-x.eu/wp-content/uploads/2022/02/Hexa-X_D3.1_v1.4.pdf.

[3] ITU-R, "Framework and overall objectives of the future development of IMT for 2030 and beyond," *DRAFT NEW RECOMMENDATION,*, 2023.

[4] 3GPP, "Study on Integrated Sensing and Communication," [Online]. Available: https://portal.3gpp.org/desktopmodules/Specifications/SpecificationDetails.aspx?specificationId=4044.

[5] Qualcomm Incorporated, "Integrated Sensing and Communications," [Online]. Available: https://www.3gpp.org/ftp/TSG_RAN/TSG_RAN/TSGR_AHs/2023_06_RAN_Rel19_WS/Docs/RWS-230175.zip.

[6] Ericsson, "Discussion on Joint Communication and Sensing in Rel-19," [Online]. Available: https://www.3gpp.org/ftp/TSG_RAN/TSG_RAN/TSGR_AHs/2023_06_RAN_Rel19_WS/Docs/RWS-230153.zip.

[7] Huawei, "Integrated Sensing and Communications in Rel-19," [Online]. Available: https://www.3gpp.org/ftp/TSG_RAN/TSG_RAN/TSGR_AHs/2023_06_RAN_Rel19_WS/Docs/RWS-230186.zip.

[8] 3GPP, "Study on channel model for frequencies from 0.5 to 100 GHz," [Online]. Available: https://portal.3gpp.org/desktopmodules/Specifications/SpecificationDetails.aspx?specificationId=3173.

[9] Fraunhofer Heinrich-Hertz-Institut, "Channel Modeling -- QuaDRiGa," [Online]. Available: https://www.hhi.fraunhofer.de/en/departments/wn/technologies-and-solutions/channel-modeling.html.

[10] IEEE, "11-21-1409-01-00bf-channel-models-for-wlan-sensing-systems," [Online]. Available: https://mentor.ieee.org/802.11/documents?is_dcn=1409&is_group=00bf&is_year=2021.

[11] NVIDIA, "Ray Tracing," [Online]. Available: https://nvlabs.github.io/sionna/api/rt.html.

[12] SIRADEL, "Volcano propagation model," [Online]. Available: https://www.siradel.com/telecommunications/volcano/.